\documentclass[12pt,twoside,a4paper]{article}
\author{Jan-Markus Schwindt}
\date{}

\title{Nothing happens in the Universe of the Everett Interpretation}

\newcommand{\ket}[1]{| #1 \rangle}
\newcommand{\bra}[1]{\langle #1 |}
\newcommand{\braket}[2]{\langle #1 | #2 \rangle}
\newcommand{\frost}{\frac{1}{\sqrt{2}}}

\begin{document} 
\maketitle 

\centerline{\small\it \quad E-mail: jan.schwindt@googlemail.com}
\vspace{0.7cm} 

\begin{abstract}
Since the scalar product is the only internal structure of a Hilbert space,
all vectors of norm 1 are equivalent, in the sense that they form a perfect
sphere in the Hilbert space, on which every vector looks the same. 
The state vector of the universe
contains no information that distinguishes it from other state vectors of the
same Hilbert space.\\
If the state vector is considered as the only fundamental entity, the world is 
completely structureless. The illusion of interacting subsystems is 
due to a ``bad" choice of factorization 
(i.e. decomposition into subsystems) of the Hilbert space.
There is always a more appropriate factorization available in which
subsystems don't interact and nothing happens at all.
This factorization absorbs the time evolution of the state vector in a 
trivial way. The Many Worlds Interpretation is therefore rather a 
No World Interpretation.\\
A state vector gets the property of ``representing a structure" only with 
respect
to an external observer who measures the state according to a specific 
factorization and basis. 
\end{abstract}

\section{Introduction}
The Everett Interpretation (EI, also known as Many Worlds)
\cite{everett,dewitt,deutsch}
is in a sense the minimal interpretation of quantum mechanics (QM):
It basically claims that only the state vector $\ket{\psi}$ of the universe
and the global Hamilton operator $H$ are fundamental. 
Everything else follows from the dynamics given by the Schr\"odinger
equation. In particular, the state vector does not represent the state {\it of}
some objects. It is the object itself.

According to the EI, there is no
mysterious collapse of the state vector, no hidden variables, no
split between a classical and a quantum realm (Heisenberg cut),
no fundamental randomness. 
A measurement is - like many other processes - a process in which a
subsystem entangles with its environment. 
The dynamics is such that the global state vector 
is (roughly speaking) split into branches, one branch for each possible
result $a$ of the measurement:
\begin{equation}\label{branching}
 \ket{\Psi_\mathrm{before}} \longrightarrow \ket{\Psi_\mathrm{after}}
 = \sum_a c_a \ket{\Psi_a}
\end{equation} 
Each branch then evolves independently,
constituting a kind of ``separate world" (hence the name Many Worlds 
Interpretation). 
The split is, however, only an {\it apparent} one; in fact the state 
vector remains a single state vector.
This is a reason why $\ket{\Psi}$,
in this context, should not be 
considered as being the state vector {\it of} some 
separately existing objects: 
These objects would {\it really} have to split when the state vector branches,
and this ``ontological excess" would make the EI unattractive.

The EI is very attractive for several reasons:
\begin{itemize}
\item It makes a minimal number of assumptions, and postulates a minimal
 number of entities (only $H$ and $\ket{\Psi}$; the branches are
 not separate entities, they are parts of $\ket{\Psi}$ and arise naturally
 through the dynamics). 
\item It demystifies the measurement process. A measurement is a
 quantum interaction like anything else. This view is supported by
 the the theory of decoherence, which shows that entanglement and
 branching are totally natural in interactions of a subsystem with its
 environment.  
\item It is closer to our traditional understanding of science, because
 of its intrinsic determinism and realism.
\end{itemize}  
The EI also has a burden, which comes with the small number of assumptions
and entities: 
It needs to explain how the world that we experience emerges from 
it, in particular
the classical behavior on macroscopic scales on one hand and the 
appearance of probabilities
in quantum measurements on the other hand.\\

The two most serious and most discussed problems of the EI are these:
\begin{itemize} 
\item How can the EI explain the observed
 probabilities in quantum measurements? I.e.,
 why is the squared norm $|c_a|^2$ of a branch equivalent to the
 probability an observer encounters for measuring the value $a$?
 If an observer performs the ``same"
 (equivalent) measurement many times, the 
 state vector branches each time, and in the end there will be 
 a branch for each combined result of the measurements. Each branch also
 contains one version of the observer.
 Each observer will conclude the probability for each value $a$
 from the statistics of the individual results he got. 
 One can show that the norm of the part of the state vector corresponding to
 branches where observers {\it don't} get the
 right probabilities converges to zero when the number of measurements is 
 increased \cite{dewitt}. 
 The remaining question is whether or not this argument solves the 
 problem (I think it does). 
 In this paper, I will not deal with the probability problem, 
 so I won't discuss this issue any further.
\item The branching in eq.(\ref{branching}) occurs only with respect to a 
 specific basis. How can we determine this basis from the dynamics? 
 This is refered to as the {\bf basis problem}. The basis problem consists
 of two parts: \\
 \\
 (A): How can the universe be split into subsystems appropriate to the 
 measurement? I.e., how can the measurement device $M$
 and the system $S$ to be 
 measured be singled out within the global $\ket{\Psi}$? In other words:
 How can we split the global Hilbert space $\mathcal{H}$ into a tensor
 product
 \begin{equation}
  \mathcal{H} = \mathcal{H_S}\otimes\mathcal{H_M}\otimes\cdots
 \end{equation}
 such that the factors represent what we see as the ``objects" of the 
 measurement? There are no ``real" objects, but the {\it apparent} objects
 must be somehow justified as properties of the evolving state vector.\\
 \\  
 (B): What is the basis of $\mathcal{H_S}$ and $\mathcal{H_M}$ along
 which the branching happens? \\
 \\
 The second part is solved by decoherence \cite{jooszeh,zurek,decobuch}: 
 {\it Once the factorization into 
 subsystems has been done}, the interactions of $S$ with the
 environment (of which $M$ is a part) 
 lead to branching according to a specific basis,
 and in each of these branches, $M$ will show a specific result.\\
 The really tough part is (A). One purpose
 of this paper is to show that it is 
 {\it the} hard problem of the EI. I will refer to it
 as the {\bf factorization problem}.  
\end{itemize}
It is interesting that the factorization problem has received so little
attention. Most authors focused on the part of the basis problem which
is solved by decoherence. Only few authors have noticed that the factorization
problem is a serious threat to the Everett interpretation
\cite{deutsch,dugic2,dugic3}.

I think the reason is mainly that in almost all situations occuring in 
practice, the subsystems are already given (particles, cats, ...).
The tensor product is considered bottom-up, from the subsystems to the 
total system, not top-down, from the total system to the deduction of an 
``appropriate" decomposition into subsystems. \\

Wallace \cite{wallace1} has 
claimed that the whole basis problem is not an issue, because
it is not necessary to specify a basis for the EI to work. The choice of 
a basis in QM is like the choice of a sequence of spatial 
``constant time" hypersurfaces in 
General Relativity (GR). It is up to the user to define such hypersurfaces
in order to solve a problem. But the physics of GR is independent of such a
choice, and GR works without the requirement of such a choice to be 
specified. 

I totally agree with Wallace that the analogy should be valid if 
the EI works. The only problem is that in GR there are frame-independent
quantities, like the Ricci scalar, or the square of the Riemann tensor.
So we can, independently of a choice of coordinates, decide whether two
given manifolds have the same properties or not.
{\bf The state vectors in QM, however, are all equivalent.} The only
structure internal to $\mathcal{H}$ is its scalar product. Taking only
the state vectors of norm 1, there is no structure available to 
distinguish them. They form a perfect complex sphere, and every vector
looks the same. This turns out to play a central 
role for the factorization problem,
and to be a challenge for the EI.
\newpage
I want to give two analogies to the argument I am presenting here:\\
\\
{\bf 1. Minkowski space:}\\
As a vacuum solution of GR, Minkowski space describes an empty universe without
curvature in which nothing happens. It is described, in an appropriate
coordinate frame, by the line element
\begin{equation}\label{minknirv}
 ds^2 = - dt^2 + dx^2 + dy^2 + dz^2
\end{equation}   
But, using a different coordinate frame, we get the line element
\begin{equation}\label{minksams}
 ds^2 = -d\tilde{t}\,^2 +\tilde{t}\,^2\left(\frac{d\tilde{r}^2}{1+\tilde{r}^2} 
 +\tilde{r}^2 d\Omega^2 \right) .
\end{equation}
This is still Minkowski space, but in this frame it looks like it
is an expanding universe with negative spatial curvature. And of 
course we can construct much worse spatial hypersurfaces, in which
Minkowski space looks like there are all kinds of spatial structures
evolving in time. 

I will call a frame like that of eq.(\ref{minknirv}) a {\bf Nirvana
frame}: a frame in which it is obvious that nothing happens. 
A frame like that of (\ref{minksams}) will be called {\bf Samsara frame}:
a frame in which it looks like something happens, although in fact nothing
happens\footnote{These terms are used as a purely metaphoric analogy
to certain ideas from Indian philosophy. I'm {\it not} trying to suggest
a real connection between physics and such a philosophy.}. 
My argument will be that in the EI, bases which  
show a branching are just Samsara frames. We can always find a Nirvana
frame which shows that in fact nothing happened at all.\\
\\
{\bf 2. Classical Phase space:}\\
In classical mechanics, Hamilton-Jacobi Theory tells us that 
for each system it is
possible to find a canonical transformation such that
the generalized position and momentum variables $Q_i$, $P_i$ are
constant. This constitutes a Nirvana frame of phase space. 
Nothing ever happens in this frame. Every classical system has ``Nirvana
character" in such a generalized phase space. If phase space
were the {\it only real thing, the fundamental stage of physics}, 
one may ask why one should ever consider coordinates
different from those of the peaceful Nirvana frame. 
But in classical mechanics,
we use this generalized version of phase space only as a {\it tool} to
describe the motion of the {\it real particles} which move through
a three-dimensional space and have {\it real} positions $q_i$ and {\it real}
momenta $p_i$, and the generalized $Q_i$, $P_i$ are useful only because
we can express $q_i$ and $p_i$ as functions of them. Alone and intrinsically
the $Q_i$, $P_i$ have no meaning and cannot have one, because they 
describe, in an appropriate frame, a universe in which nothing happens at all.
I will argue that the same is true for the state vector $\ket{\Psi}$ of
the universe.\\

Section 2 contains an outline of the EI, and a list of things that need to be
explained from the dynamics of $\ket{\Psi}$ if it is really fundamental.
The factorization problem and its implications are discussed in detail.\\
In section 3 I will argue that $\ket{\Psi}$ {\it per se} 
(without the help of distinguished operators acting on distinguished subspaces) 
contains no information
at all, and that it is easy to choose a basis where $\ket{\Psi}$ describes
a structureless universe in which nothing happens. \\
In section 4,
a simple measurement process is discussed. It is shown that the branching 
explanation of the process via entanglement and decoherence works only 
{\it bottom up} but not {\it top down}. I.e., if a factorization of
the Hilbert space is {\bf already given}, 
we can describe the interactions within
that factorization (i.e. between the chosen subsystems)
and get a branching that happens in the
full Hilbert space according to a preferred basis {\it within that 
factorization}, the ``decoherence basis". But if one starts from the full
Hilbert space and the dynamics of the full state vector, it is possible
to find a different factorization 
in which there are {\it no interactions at all} 
between the new subsystems. \\
The conclusion (section 5) is that if $\ket{\Psi}$
were truely fundamental, choosing a basis in which the world has a structure 
is like choosing
a coordinate system in which Minkowski space is an expanding universe. To
render QM a senseful theory, one has to provide a distinguished factorization
of the Hilbert space,
based on entities other than $\ket{\Psi}$. Or one has to give up the idea
of a universal $\ket{\Psi}$ altogether. 
Possibilities and consequences are discussed.\\
Finally (section 6), the relation of the argument to other
work in physics and philosophy is discussed.

\section{The Everett Interpretation and the factorization problem}
The postulates of QM provided by a typical textbook are:
\begin{enumerate}
\item The state of a quantum system $S$ is described by ray of vectors in
 a complex Hilbert space $\mathcal{H}$. Usually one represents the 
 state by a normalized element $\ket{\Psi}$ of the ray
 (called the state vector), $\braket{\Psi}{\Psi}=1$. 
\item To each physical observable belonging to $S$ corresponds a hermitean
 operator $A$ acting on $\mathcal{H}$. In a measurement
 \begin{enumerate}
  \item the measured value $\lambda$ is an eigenvalue of $A$; 
  \item the state $\ket{\psi}$ of the system 
   gets projected onto the eigenspace corresponding
   to $\lambda$ (``collapse"); 
  \item the probability for obtaining $\lambda$ is 
   $\braket{\psi}{P_\lambda\psi}$, where $P_\lambda$ is the projection on 
   the eigenspace.
 \end{enumerate} 
\item The time evolution of a state $\ket{\psi}$ is determined by a
 Schr\"odinger equation
 \begin{equation}
  i\frac{d}{dt}\ket{\psi}=H\ket{\psi}
 \end{equation}
 where $H$ is the Hamilton operator representing the observable which measures 
 the system's energy.
\end{enumerate}
The claim of the EI is essentially
that only postulates 1 and 3 are fundamental, while postulate 2 arises 
as a consequence of 1 and 3, on
a subjective level, i.e. as the impression gained by an observer performing 
a series of measurements.\\

The description of a measurement given by the EI is 
claimed to be as follows:
A measurement is a QM process that is completely covered by the dynamical
evolution of the state vector  $\ket{\Psi}$
of system + measurement device + observer 
+ environment, according to the Schr\"odinger equation. (The Hilbert space
in which this state vector lives is the tensor product of the Hilbert spaces
corresponding to each of the items). The Hamilton operator contains interaction
terms which enforce an entanglement of the system with the measurement
device, and further with the observer and the environment. As a consequence, 
$\ket{\Psi}$ is (approximately, in an appropriate basis)
split into several branches, i.e. a sum of terms, where each
term corresponds to one possible result of the measurement. Each term 
contains the state of an observer who thinks he has projected the state
of the system according to a hermitean operator. But in fact, no 
projection has taken place, only an entanglement, and the only
hermitean operator acting was the global Hamilton operator. \\

There are a number of things the EI has to explain if it is 
supposed to be the correct stage for QM:
\begin{itemize}
\item {\bf Factorization}:\\
How does the appearance of {\bf objects} arise? I.e., why
is it ``natural" to factorize the global Hilbert space into a tensor
product of Hilbert spaces, such that each factor makes
sense to us as the Hilbert space of an object we can distinguish from 
the rest of the ``world", where the ``world" is an arrangement of many
such objects. This is the factorization problem mentioned in the Introduction.
Given that there are infinitely many possibilities to factorize a Hilbert
space $\mathcal{H}$ (unless the dimension of $\mathcal{H}$ is a prime number):
Why is there a choice that is distinguished by
describing the ``objects" of our ``world"? The preferred choice must somehow
arise from the state vector and its dynamics. It must be a choice that
lets the state vector or its time evolution 
appear in a particularly simple way, or
that lets the dynamics of the chosen subsystems appear as independent as 
possible, or something alike.\\ \\
Tensor factorizations can be constructed in the following way: If the 
dimension of $\mathcal{H}$ is a finite number $n=p q$, choose a basis
and label the basis vectors with double indices, $\{\ket{e_{ij}}\}$,
where $i$ runs from 1 to $p$, $j$ from 1 to $q$. Then write
\begin{equation}\label{factorize}
 \ket{e_{ij}} = \ket{f_i} \otimes \ket{g_j}
\end{equation}   
and consider the $\{\ket{f_{i}}\}$ as a basis of subsystem S1, the 
$\{\ket{g_{j}}\}$ as a basis of subsystem S2. This {\it defines} the 
subsystems. If $p$ or $q$ is non-prime, the factorization can be continued.
If the dimension of $\mathcal{H}$ is infinite, one or both of the indices
can be chosen to run from 1 to $\infty$. Some choices
$\{\ket{e^{(1)}_{ij}}\}$ and $\{\ket{e^{(2)}_{ij}}\}$ are equivalent in the
sense that the transformation between them is only a transformation
within S1 and S2 separately, i.e. they both correspond to the same split
of the entire system into two subsystems, only the basis for those subsystems
is different. But for most choices, the split into an S1 and an S2 will be 
truely different. I will refer to $\{\ket{e_{ij}}\}$ as the {\bf factorization
basis}.\\
\\
{\bf Are we allowed to choose the factorization in a time-dependent way?} 
With time-dependent $\{\ket{e_{ij}(t)}\}$? Since $\ket{\Psi}$ depends on time,
and we are discussing subsystems that arise 
as ``meaningful" from its time dependent
behavior, why should we not allow these emerging subsystems to move their
``position"/``orientation" within the global Hilbert space? 
(With subsystems I mean the Hilbert space factors, not the 
corresponding factors of the
state vector.)
This is also
reasonable if one wants to keep a certain equivalence between the {\bf pictures}
of QM. A picture is a {\bf vector identification map (VIM)} that tells
us which state vector in the Hilbert space at time $t_2$ is ``equal" to 
a state vector in the Hilbert space at time $t_1$. In the Schr\"odinger
picture, the state vector evolves according to the Schr\"odinger equation; 
in the Heisenberg picture, it does not evolve at all, i.e. the state vectors
are {\it identified} along their trajectory. If a factorization is constant
in the Schr\"odinger picture, it will in general not be constant in the 
Heisenberg picture, and vice versa. Deutsch \cite{deutsch}, for 
example, in his approach to the EI,
discusses everything in the Heisenberg picture. His factorizations
have to depend on time, because otherwise no entanglement could happen.
The state vector is constant, so the only way how it can represent 
the process of entanglement between
interacting subsystems is when its decomposition is {\it not} constant.
If we want to keep the pictures on equal footing, we should also allow
for time dependent decompositions in the Schr\"odinger picture.
However, one may argue that in the EI, the Schr\"odinger picture is
singled out: The dynamics should happen in the state, not in the operators,
because operators are not fundamental \cite{decobuch}. Still, even when
the Schr\"odinger picture is preferred, I don't see any principle that forbids
subsystems to change their ``position"/``orientation" in the global Hilbert
space. \\
\\
Some comments about the time evolution of the subsystems' states are in order.
The Hamilton operator (here for the generic case 
of an explicit time dependence)
can, for a given split into two subsystems, be decomposed
\begin{equation}\label{hdecomp}
 H(t) = H_1(t) \otimes \mathbf{1} + \mathbf{1} \otimes H_2(t) 
 + H_\mathrm{int}(t),
\end{equation} 
and then, because there are infinitely many ways to write $H(t)$ in such a form,
one may choose the one in which $H_\mathrm{int}(t)$ is,
at each time, ``minimal" in a certain sense
I will not specify any further.
Again, this is an unusual task, because usually the subsystems, their
internal Hamilton operators and their interaction are already given.
Here, the decomposition of $H(t)$ {\it defines} the interaction.\\
If (a) the factorization is constant, i.e. $\{\ket{e_{ij}}\}$ does not
depend on time, and if (b)
$H_\mathrm{int}(t)\ket{\Psi(t)}$ is zero for
some time interval, 
and if (c) $\ket{\Psi(t)}=\ket{\psi_1(t)}\otimes\ket{\psi_2(t)}$
at the beginning of that time interval, then 
each subsystem state will evolve 
independently according to a Schr\"odinger
equation with Hamilton operators $H_1(t)$ and $H_2(t)$, respectively.\\
If the factorization is {\it not} constant, i.e. $\{\ket{e_{ij}(t)}\}$ does
depend on time, there may still be time intervals in which the subsystem
states evolve independently. The evolution is still unitary, so each 
subsystem state
will appear under the effect of a Schr\"odinger equation. The corresponding
``apparent" Hamilton operators, however, {\bf are not the operators
$H_1$ and $H_2$ of the decomposition} (\ref{hdecomp}) {\bf any more}.
Part of the time dependence was brought in ``by hand", via the time
dependence of the factorization, and that part 
cannot be derived from the global Hamilton operator $H(t)$.
\item {\bf Space}:\\
How does the appearance of {\it space} arise? The state vector
lives in an infinite or very high dimensional complex Hilbert space. Why
do the ``objects" look like they are layed out in a three-dimensional 
real space? Sure, the state $\ket{\Psi}$ was constructed {\it as} a wave
function $\Psi(x_1,\cdots,x_{3n})$ (if we restrict ourselves to the
QM of $n$ scalar particles moving in three dimensions for a moment);
and the Hamilton operator $H$ was contructed in this spatial basis too,
and is, with respect to its symmetries, most clearly written 
in the spatial basis, i.e. written {\it as} an operator acting on wave 
functions. The state vector will in general not have the
same symmetries as $H$, just as in general a solution to some equations
of motion doesn't have the symmetries of the equations.
(The symmetries of the equations are reflected as symmetries between
different solutions, not as symmetries within one solution.) 
Basically $\ket{\Psi}$ is an abstract vector that appears {\it as} 
a wave function if we write it in a spatial basis. But why does an
observer, who arises as a certain part of the state vector, see his
``world" of ``objects" layed out in a space which corresponds to this
specific basis?
(Here, ``part" means something like a factor in  
a certain branch of $\ket{\Psi}$, possibly after ``tracing out"
practically inaccessible degrees of freedom of the environment.
I will not discuss the possible complications in defining an observer
in the EI, because it is not the main concern here. For example,
``tracing out" implies that one considers a density operator 
instead of a state vector, i.e. the observer is involved in 
some kind of ensemble picture, which I want to avoid here.) 
\\ \\
In fact, there are infinitely many different spaces that can serve
as a basis. (When I say a space ``serves as a basis" I mean that all
vectors of the Hilbert space are written as wave functions in $n$ copies
of that space, where $n$ is the number of particles). Consider the
case $n=1$, and two orthogonal state vectors $\ket{e_1}$ and $\ket{e_2}$
of norm 1.
Written as wave functions in some ``$x$-space"
(in which $H$ takes a particularly simple and symmetric form), we may have
\begin{equation}
 \braket{\mathbf{x}}{e_1}= f_1(x), \quad \braket{\mathbf{x}}{e_2}= f_2(x)
\end{equation}
where $f_{1,2}$ are specific functions which are orthogonal to each other.
We may define a new space, ``$y$-space" via
\begin{equation}\label{nsone}
 \braket{\mathbf{y}}{e_1}= f_2(y), \quad \braket{\mathbf{y}}{e_2}= f_1(y),
\end{equation}
and for all vectors $\ket{e_i}$ which are orthogonal to $\ket{e_1}$ and 
$\ket{e_2}$, we define
\begin{equation}\label{nstwo}
 \braket{\mathbf{y}}{e_i} = \braket{\mathbf{x}=\mathbf{y}}{e_i}, 
\end{equation}
where $\bra{\mathbf{x}=\mathbf{y}}$ is the (pseudo-)vector for which
$\mathbf{x}$ has the same value as $\mathbf{y}$ on the left hand side
of the equation.
That is, in $y$-space the roles of $\ket{e_1}$ and $\ket{e_2}$ are exchanged.
In terms of $x$-space, the vector $\ket{e_1}$ represents the function $f_1$;
in terms of $y$-space, the same vector represents the function $f_2$, and 
vice versa for $\ket{e_2}$. For all vectors orthogonal to  
$\ket{e_1}$ and $\ket{e_2}$, the corresponding functions are the same in
both spaces. 
This is always possible because there must be a unitary transformation
which exchanges $\ket{e_1}$ and $\ket{e_2}$ and leaves all vectors
orthogonal to them invariant.
The relations (\ref{nsone}) and (\ref{nstwo}) {\it define} the new space.
The two spaces are totally different. You don't get $y$-space by simply
moving around some points of $x$-space. And yet we can write $\ket{\Psi}$ as
a wave function in $x$-space or in $y$-space, and there is no reason why
$\ket{\Psi}$ should look simpler in $x$-space than in $y$-space. 
(This is well known for $x$-space versus $k$-space, the Fourier transformed
space. This is just a reminder that there is an infinity of such spaces.)
The Hamilton operator will look very unpleasant in $y$-space. The position
operator $X$ looks nice in $x$-space and unpleasant in $y$-space. The
position operator $Y$ looks nice in $y$-space and unpleasant in $x$-space, etc.

So, the question is: Why do we appear to live in the space in which the
Hamilton operator looks particularly simple? Does the time evolution
of $\ket{\Psi}$, induced by this Hamilton operator, also have particularly nice properties if expressed in that space?
(We will see that this is not the case.)\\
\\
In quantum field theory (QFT) things are more complicated, but not 
qualitatively different. The Hamilton operator is construcuted as an integral
over operators (operator-valued distributions) which are local with respect to
some $x$-space. So, $x$-space takes part in the {\it construction} of $H$.
But we can refer to $H$ as an abstract hermitean operator acting on the 
global Hilbert space, without refering to how it was constructed. Again,
$H$ may take on its simplest form 
if it is written in terms of such an integral. But again the question is
why an observer, arising as some part of the global state vector, 
``sees" this particular space over which the integrals run.
\\ \\
The question of how space arises can be considered a part of the 
factorization problem. If the distinguished factorization is such that the 
Hamilton operators $H_i$ of the subsystems $S_i$ still have their simplest
form if written in $x$-space (i.e. as operators acting on wave functions
in $x$-space), and the interaction Hamiltonian $H_\mathrm{int}$ between
the subsystems describes interactions {\it local} in $x$-space, then we have
a good reason to consider $x$-space as distinguished. \\
There are more complications, for example how to identify the $x$-spaces of
the subsystems with each other. 
In what sense can we say that the $x$-space in which 
$S_1$ is expressed is the ``same" space as the $x$-space in which $S_2$
is expressed? 
Intuitively it seems reasonable that the answer is somehow contained in the
local interaction again. The interaction gets strong when the $x$-value of
$S_1$ gets close to the $x$-value of $S_2$ (if both $S_1$ and $S_2$ are 
in states such that their wave functions are narrow in $x$-space,
so that we can speak of an ``$x$-value of $S_{1,2}$").
In that sense we may say that $S_1$ gets ``close" to $S_2$.
Again, I will not go any deeper here.  

\item {\bf Classicality}:\\
Why do the ``objects", in particular if they are macroscopic,
appear to have classical properties? I.e., why don't we see any 
superpositions, objects smeared out in space etc? These questions
can be asked only if the previous issues have been resolved, so
we know what the objects are and what space is. But then these
particular questions are resolved by decoherence. The interactions
carry the superpositions of an object into the environment via
entanglement, such that the superposition is, after a short time
of interaction, a global superposition, i.e. a branching structure
of the global state vector, instead of a local superposition 
just within the object. Within a branch, the object appears to have
classical properties, in the sense that the properties which
are distinguishable through the interactions have definite values
within a branch. 
\item {\bf Measurement}:\\
How does the second postulate of QM emerge from 1 and 3?
This is also resolved by decoherence. A measurement is a controlled interaction
between a measuring device $M$ and a system $S$. The interaction is such that
a certain property of $S$ is distinguished, i.e. the different components of 
$\ket{\Psi_S}$ with respect to a specific basis of $\mathcal{H}_S$ have a
different impact on $M$. This leads to a branching of the global state vector
where in each branch only one component of $\ket{\Psi_S}$ survives.
Together with the value shown by the pointer of $M$ this gives an observer
(who lives in one branch) the impression he has projected $\ket{\Psi_S}$
with a specific hermitean operator. The appearance of probabilities is still
under discussion, but as mentioned in the introduction a reasonable 
solution exists.
\end{itemize}
From the discussion in this section we may conclude that the EI is 
a reasonable and elegant framework for QM {\bf as soon as we can solve
the factorization problem}. 

\section{The state vector of the universe}
The factorization problem with all its implications, as described in the
previous section, looks quite complicated. The solution is remarkably
simple: Nothing of all this happens in the EI.\\

We have only two fundamental entities at hand: the state vector of 
the universe $\ket{\Psi}$, and the Hamilton operator $H$. 
We assume $H$ to be not explicitly time-dependent, because a time-dependent
$H$ is hard to reconcile with the symmetries of special or general relativity,
or with a global energy conservation.
We may forget about how $H$ was constructed and consider it just as an
abstract hermitean operator acting on $\mathcal{H}$, the space in which
$\ket{\Psi}$ lives. The time evolution of $\ket{\Psi}$ takes the 
simplest form in the basis where $H$ is diagonal,
\begin{equation}\label{peace}
 \ket{\Psi(t)} = \sum c_n e^{-iE_nt} \ket{n},
\end{equation}  
where $\ket{n}$ are the eigenvectors and $E_n$ the corresponding eigenvalues
(for simplicity we assume a discrete spectrum).
Each component of $\ket{\Psi}$ peacefully performs its phase rotations. 
In this form, it seems hard to believe that $\ket{\Psi}$ describes
interacting subsystems. Yes, $H$ was constructed as the 
operator of ``energy" present in some interacting particles or fields. But
how are these interactions represented in $\ket{\Psi}$? Not at all.
At least not when we look at it in this simple form.\\

Now let's discuss factorizations into subsystems. One may ask
why the ``peaceful" state (\ref{peace}) should be factorized at all. 
Let's ignore the why-question and just do it. However, when we do 
it, we should look for factorizations as simple as possible. If we find
a ``Nirvana factorization", according to which nothing happens, we
should prefer it as compared
to a ``Samsara factorization", according to which it seems
like something happens; just as the Minkowski metric (\ref{minknirv})
should be 
considered more ``appropriate" than (\ref{minksams}). That is because
a Samsara frame constitutes an {\it arbitrary} and {\it unnecessary}
complication. Of course, both metrics (\ref{minknirv}) and
(\ref{minksams}) are {\it correct}. But if we ask: ``Is Minkowski space
associated with the prediction of an expanding universe?", 
the answer should be no. Similarly, if we find a Nirvana frame for 
$\ket{\Psi}$, we can no longer claim that the EI predicts a universe
in which subsystems interact and entangle with each other.\\

I will divide the following discussion into two parts: At first I will
consider only time-independent factorizations of Hilbert space, then
also time-dependent ones. As I discussed in the previous section, I
think the second option is fine, but some people may be more
conservative.\\

If a factorization has to be {\bf time-independent}, one can proceed in two
steps: At first one may look for a decomposition in which the subsystems
don't interact at all, i.e. the Hamilton operator can be written as
\begin{equation}
 H = H_1 \otimes \mathbf{1} \otimes \mathbf{1} + \mathbf{1} \otimes H_2
 \otimes \mathbf{1} +  \mathbf{1} \otimes  \mathbf{1} \otimes H_3 + \cdots
\end{equation}
This is the case if all eigenvalues $E_n$ can be written in the form
\begin{equation}
 E_n = E^{(1)}_i + E^{(2)}_j + E^{(3)}_k + \cdots
\end{equation}
for some index sets $I$, $J$, etc., $i\in I$, $j\in J$, etc., 
such that all combinations $E^{(1)}_i + E^{(2)}_j + E^{(3)}_k + \cdots$ 
occur in the spectrum of $H$, with the correct degeneracies. 
(If the dimension of $\mathcal{H}$ is a finite number $d$, the product of
the cardinalities of $I$, $J$, etc. must be $d$.) 
Then we may relabel
\begin{equation}\label{multiind}
 \ket{n} = \ket{ijk\cdots},
\end{equation}
take this as the factorization basis,
\begin{equation}\label{ijknirv}
 \ket{ijk\cdots} = \ket{i}\otimes\ket{j}\otimes\ket{k}\cdots,
\end{equation}
and interpret the $E^{(\alpha)}_i$ as the energy eigenvalues of the
subsystems $S_\alpha$. This is a Nirvana factorization, there is no interaction
between the subsystems, and nothing happens except phase rotations for
each energy component, in each subsystem independently.\\

If such a decomposition does not exist, we may {\it still} relabel the 
$\{\ket{n}\}$ in multiindex form, as in (\ref{multiind}), but this
time in an arbitrary way (consistent with the dimension of $\mathcal{H}$,
if that is finite), and use this as the factorization basis,
as in (\ref{ijknirv}). Now, there are interactions between the such
defined subsystems. But their only effect is to cause different speeds for
the phase rotations of the branches: The time evolution of $\ket{\Psi}$ reads
now
\begin{equation}
 \ket{\Psi(t)}=\sum c_{ij\cdots} e^{-iE_{n(ij\cdots)}t}\ket{i}\otimes\ket{j}
 \otimes \cdots
\end{equation} 
Each term in this sum can be interpreted as a branch of the world,
according to the chosen factorization. But all branches are there
right from the beginning, they are not due to some entanglement that
happens in a physical process. Each branch peacefully rotates with its
frequency $E_{n(ij\cdots)}$, and nothing else happens. Therefore, 
we can still consider this a Nirvana frame. \\

If one allows {\bf time-dependent factorizations} (and I think this is what one
should do), the situation becomes even simpler. 
Now we don't even have to assume that $H$ is time-independent, 
and we don't need to work with a discrete spectrum to keep things simple.
We may just choose a
time-dependent basis vector $\ket{e_1(t)}$ which absorbs the time evolution
of $\ket{\Psi(t)}$,
\begin{eqnarray}
 \ket{e_1(t)} = \ket{\Psi(t)}.
\end{eqnarray}
This means, we effectively move to the Heisenberg picture. 
If we take $\ket{e_1(t)}$ as part of a factorization
basis (the rest of the basis is irrelevant), i.e.
\begin{equation}\label{absnirv}
 \ket{e_1(t)} = \ket{f_1}\otimes\ket{g_1}\otimes\cdots,
\end{equation}
we have a real Nirvana factorization: In the subsystems
according to this factorization, nothing
happens at all, not even phase rotations. 
Each subsystem is in an eigenstate with respect to its 
``apparent" Hamilton
operator, with eigenvalue zero. There are no branches
and no interactions,
$\ket{\Psi(t)}$ is once and for all given by the tensor product
(\ref{absnirv}).\\

This simple choice is possible due to two facts: (a) All state vectors 
of norm 1 are equivalent, there is no structure available 
in the Hilbert space which distinguishes $\ket{\Psi}$ from any other
state vector. It makes a difference with respect to $H$, but who
cares? Nobody is ever going to make a measurement on $\ket{\Psi}$
using $H$ (measure the ``energy of the universe"). 
Instead, the EI assumes that everything follows from 
``inside" $\ket{\Psi}$ and its time evolution. Yes, different
$\ket{\Psi}$s have different time evolutions. But: (b) Since we
have no other
guiding principle how to choose subsystems of $\mathcal{H}$,
we can just choose them such that they absorb this time evolution 
for the {\it given} $\ket{\Psi}$ in the
most trivial way.\\

This is a fundamental property of Quantum Mechanics: The state vector 
{\it per se} does not contain any information at all. The information
arises only when a state is measured using a specific operator. 
The presence of ``information" or ``structure" requires both, 
the state vector and the operator. 
Here, ``measured" can be understood either in 
the ``mystical" sense of the Copenhagen Interpretation, or in the sense
of entanglement + decoherence. 
In any case, there must be an environment to which the 
state vector really represents a state. For the state vector
of the universe, there is no such environment.\\

The tensor product structure of Hilbert space is usually only considered
bottom-up, not top-down. That is, one starts from a given product structure
(system, environment) and given interactions, and computes the time evolution
of this product structure. The combined state vector is always written
{\it as} a vector in a tensor product of the subsystem Hilbert spaces.
But for the EI to make sense, we have to go the other way: we have to start
from the combined Hilbert space and factorize it in a meaningful way such
that interacting subsystems emerge. In this section I have demonstrated
that this does not work, or more precisely:
We {\it can} factorize it into interacting subsystem,
but the way to do this is unnecessarily
complicated and completely arbitrary. 
In the next section I will discuss the bottom-up
versus top-down problem by the means of a simple example. 

\section{The measurement process bottom-up and top-down}
We base our discussion on an idealized simple measurement process. 
Consider
a system $S2$ consisting of a measurement device $M$ and an object 
to be measured, $S1$, 
with corresponding Hilbert space
\begin{equation}\label{tensfaca}
 \mathcal{H}_{S2} = \mathcal{H}_M\otimes\mathcal{H}_{S1}.
\end{equation}
For simplicity, we take $\mathcal{H}_{S1}$ to be two-dimensional
and $\mathcal{H}_M$ three-dimensional. 
We denote the basis vectors of $\mathcal{H}_{S1}$ in which the measurement is
diagonal $\ket{\uparrow}$ and $\ket{\downarrow}$. 
Before the measurement,
$M$ is in the state $\ket{0}$. After the measurement,
$M$ will be in the state $\ket{+}$ if $S1$ is in the state 
$\ket{\uparrow}$, or $\ket{-}$ if $S1$ is in the state $\ket{\downarrow}$.
Here, $\ket{0}$, $\ket{+}$ and $\ket{-}$ are orthogonal and 
can be thought of as pointer positions of the measurement device:
$\ket{+}$ indicates that $\ket{\uparrow}$ has been measured, 
$\ket{-}$ indicates that $\ket{\downarrow}$ has been measured.

According to the Everett interpretation, the whole measurement process
can be described by the unitary evolution induced by the Hamilton
operator $H$ acting on the six-dimensional space $\mathcal{H}_{S2}$. That is,
$H$ contains an interaction term between $M$ and $S1$ which lets an initial
state $\ket{0}\otimes\ket{\uparrow}$ evolve into 
$\ket{+}\otimes\ket{\uparrow}$, and similarly for $\ket{\downarrow}$:
\begin{eqnarray}
 \ket{\Psi}_\mathrm{before} = \ket{0}\otimes\ket{\uparrow} &\longrightarrow&
 \ket{\Psi}_\mathrm{after} = \ket{+}\otimes\ket{\uparrow} \\
 \ket{\Psi}_\mathrm{before} = \ket{0}\otimes\ket{\downarrow} &\longrightarrow&
 \ket{\Psi}_\mathrm{after} = \ket{-}\otimes\ket{\downarrow}
\end{eqnarray}
For simplicity we assume that 
\begin{itemize}
\item the interaction is switched on at a time
  $t_\mathrm{before}$ and switched off at $t_\mathrm{after}$
\item $\ket{\Psi}_\mathrm{before}$ is constant before $t_\mathrm{before}$
\item $\ket{\Psi}_\mathrm{after}$ is constant after $t_\mathrm{after}$
\item The interaction is constant between $t_\mathrm{before}$ and
  $t_\mathrm{after}$, so $\ket{\Psi}_\mathrm{before}$ gets
  smoothly rotated into $\ket{\Psi}_\mathrm{after}$.
\end{itemize}
If we ascribe values to the pointer states $\ket{+}$ and $\ket{-}$, say $+1$
and $-1$, then this time evolution generates the picture that $M$ has
``acted on $S1$" with a hermitean operator which in the chosen basis of
$\mathcal{H}_{S1}$ has the matrix form $\sigma_z = 
\left(\begin{array}{cc}1&0\\0&-1\end{array}\right)$. This operator is a 
derived property of the process.

The superposition principle tells us that for an initial $S1$-state
$\frost(\ket{\uparrow}+\ket{\downarrow})$ the time evolution is
\begin{equation}\label{entangl}
  \ket{\Psi}_\mathrm{before} = 
   \frost\ket{0}\otimes(\ket{\uparrow}+\ket{\downarrow})
  \longrightarrow
  \ket{\Psi}_\mathrm{after} = \frost ( \ket{+}\otimes\ket{\uparrow}
   + \ket{-}\otimes\ket{\downarrow})
\end{equation}
Written in this form, the time evolution of $\ket{\Psi}$ tells a
{\it Samsara story}: The object $S1$ entangles with the measurement 
device, thereby splitting $\ket{\Psi}$ into two branches; one branch
with measurement result $+1$, one with result $-1$.\\

We may choose the following orthonormal basis for $\mathcal{H}_{S2}$:
\begin{eqnarray}\label{tensbas}
 \ket{e_1}&=&\ket{0}\otimes\ket{\uparrow},\quad 
 \ket{e_2}=\ket{0}\otimes\ket{\downarrow},\quad 
 \ket{e_3}=\ket{+}\otimes\ket{\uparrow},\\
 \ket{e_4}&=&\ket{+}\otimes\ket{\downarrow},\quad 
 \ket{e_5}=\ket{-}\otimes\ket{\uparrow},\quad 
 \ket{e_6}=\ket{-}\otimes\ket{\downarrow}. 
\end{eqnarray}
Written in components according to this basis, 
the evolution given in (\ref{entangl})
becomes
\begin{equation}
 (\frost,\frost,0,0,0,0) \longrightarrow (0,0,\frost,0,0,\frost)
\end{equation}
Now we define
\begin{eqnarray}
 \ket{e'_1} &=& \frac{1}{2}(\ket{e_1}+\ket{e_2}+\ket{e_3}+\ket{e_6})\\
 \ket{e'_2} &=& \frac{1}{2}(\ket{e_1}+\ket{e_2}-\ket{e_3}-\ket{e_6})
\end{eqnarray}
These two vectors can be completed
by four other vectors 
to form another orthonormal basis of $\mathcal{H}_{S2}$. 
(The choice of the four other basis vectors is arbitrary, 
as long as they are orthonormal to the first two, because 
$\ket{\Psi}$ has no components in their direction.) 
In this new
basis, the time evolution in (\ref{entangl}) becomes
\begin{equation}\label{snirvana}
 (\frost,\frost,0,0,0,0) \longrightarrow (\frost,-\frost,0,0,0,0).
\end{equation} 
The story told by this encryption of $\ket{\Psi}$ is a {\it Nirvana story}:
$\ket{\Psi}$ is a sum of two components. One of them is unchanged after the
process, the other one acquires only a phase exp($i \pi$). There is 
no sign of any interaction or entanglement in this basis. \\

What happened? With the new basis we left the original picture of the
six-dimensional Hilbert space $\mathcal{H}_{S2}$ being obtained from 
a tensor product. We considered it a space on its own right, without
reference to the tensor decomposition $\mathcal{H}_M\otimes\mathcal{H}_{S1}$
it was constructed from. This is in accordance with the Everett interpretation,
where the larger Hilbert space is always more fundamental, and a
meaningful tensor factorization should emerge from the dynamics in that
larger Hilbert space. Then we constructed a basis in which the 
unitary time
evolution operator $U(t_\mathrm{after},t_\mathrm{before})$ is diagonal.
In this
basis the time evolution consists only of phase rotations of each
component.\\

There are infinitely many possibilities to factorize a six-dimensional Hilbert
space into the tensor product of a two- and a three-dimensional Hilbert space.
Each six-dimensional basis can serve as a starting point to factorize the
basis vectors similarly as in (\ref{tensbas}). For example, if we use 
$\ket{e'_1}$, $\ket{e'_2},\cdots$ as a starting point, we may factorize
\begin{equation}\label{tensfacb}
 \mathcal{H}_{S2} = \mathcal{H}_{M'}\otimes\mathcal{H}_{S1'}
\end{equation} 
by writing 
\begin{equation}
 \ket{e'_1}=\ket{0'}\otimes\ket{\uparrow '},\quad 
 \ket{e'_2}=\ket{0'}\otimes\ket{\downarrow '}
\end{equation}
etc., which turns the evolution (\ref{entangl}) into
\begin{equation}
 \frost\ket{0'}\otimes(\ket{\uparrow '}+\ket{\downarrow '}) \longrightarrow 
 \frost\ket{0'}\otimes(\ket{\uparrow '}-\ket{\downarrow '})
\end{equation}
No entanglement, no measurement 
takes place in this decomposition; the phase rotation of the 
$\ket{\downarrow '}$ component is internal to $S2'$ and can be considered
as being due to an $S2'$-internal Hamilton operator $H_{S2'}$.  
And yet it is 
exactly the same time evolution as in (\ref{entangl}), only written in a
different factorization of $\mathcal{H}_{S2}$.\\

The ``Nirvana factorization" (\ref{tensfacb}) tells a simpler story than
the ``Samsara factorization" (\ref{tensfaca}).
Why would one still choose the factorization (\ref{tensfaca}) over 
(\ref{tensfacb})? The reason is, of course, that an external observer $O$
may be able to ``read off" $M$, but not $M'$ (fortunately, since $M'$
wouldn't tell him anything but ``Nothing happened"). 
That is, he makes himself
a ``measurement" (observation)
which operates on $\mathcal{H}_M$, not on  
$\mathcal{H}_{M'}$. The way in which the observer interacts with 
the system $S2$ sets a 
preference for
\begin{itemize}
\item the tensor decomposition (\ref{tensfaca}), i.e. he 
 sees the measurement device $M$ as a distinct object;  
\item the choice of basis for $\mathcal{H}_M$, i.e. he sees
 the state of $M$ as a choice between specific pointer positions. 
\end{itemize}
Only due to this specific interaction with an observer {\it outside}
of $S2$  the Samsara story of a measurement taking place becomes
meaningful.\\

The problem is: We can play the same game again.
According to the Everett interpretation, this new interaction 
can again be described by the unitatry evolution of a state vector
in a larger Hilbert space $\mathcal{H}_{S3}$, 
which is the tensor product
of $\mathcal{H}_{S2}$ and the Hilbert space $\mathcal{H}_{O}$ 
in which the observer ``lives",
\begin{equation}
 \mathcal{H}_{S3} = \mathcal{H}_{O} \otimes \mathcal{H}_{S2}.
\end{equation}
The story given by this description is that of an observer entangling
with the system $S2$, splitting into a branch where the
observer sees the measurement device in position $\ket{+}$ and another
branch where he sees the measurement device in position $\ket{-}$.

Again, however,
there is a basis of the combined Hilbert space $\mathcal{H}_{S3}$
in which the unitary time evolution is diagonal. 
In this basis, nothing happens apart from phase rotations of some components. 
Applying the same argument as before,
we see that the ``Samsara factorization" into an observer and the System $S2$
is a bad choice, because the according tensor 
split of the Hilbert space tells an unnecessarily complicated story.\\

We can go on like this, arguing that the system $S3$ again interacts with its
environment, in particular with other observers, who see the first 
observer $O$ as a separate entity, 
thereby favoring the original factorization.
But at some point, we reach the state vector of the universe. 
Now there is no external interaction, no external observer anymore. 
At this point, there is no valid argument anymore why a Samsara factorization
should be preferred, and the arguments of section 3 apply.\\

Some remarks:\\
1. A Nirvana factorization of $S3$ can be obtained in analogy with $S2$:
Let $\ket{\Phi}_\mathrm{before}$ and $\ket{\Phi}_\mathrm{after}$ be the 
$S3$-states before and after both the measurement has been performed {\it and} 
the observer has read off the result. Then define
\begin{equation}
 \ket{\alpha} = \frost(\ket{\Phi}_\mathrm{before}+\ket{\Phi}_\mathrm{after}),
 \quad 
 \ket{\beta} = \frost(\ket{\Phi}_\mathrm{before}-\ket{\Phi}_\mathrm{after}).
\end{equation}
Since $\ket{\Phi}_\mathrm{after}$ is orthogonal to $\ket{\Phi}_\mathrm{before}$
(the state of $M$ after the measurement has no overlap with the state before
the measurement), we can take $\ket{\alpha}$ and $\ket{\beta}$ as part of 
an orthonormal basis, and in this basis the time evolution is exactly as in 
(\ref{snirvana}), and we may factorize $S3$ according to this basis.

In a less idealized measurement, without a definite $t_\mathrm{before/after}$
where the interaction is turned on/off, and without perfectly orthogonal
$\ket{\Phi}_\mathrm{before/after}$, this way to define a Nirvana
factorization does not work anymore. Instead, one has to follow the
methods outlined in section 3.\\
\\
2. Deutsch \cite{deutsch} 
has noticed the factorization problem and suggested a solution.
He argued that {\it if} a measurement has taken place, there is a 
unique factorization in which certain nice properties hold for the subsystems
$M$ and $S1$; in particular, the time evolution looks like (\ref{entangl}) 
in this
factorization. Since there is only one such factorization, the factorization
problem is solved, he claims. 
The flaws in his argument have already been demonstrated on a rather technical level in \cite{fosterbrown}. 
Now we see there is a really fundamental problem with his argument: 
{\bf The mere
fact that a measurement (or an interaction)
has taken place is already a property of the 
factorization, not of the state evolution.} Deutsch erroneously 
considered the 
happening of a measurement a frame-independent property of the state evolution,
and therefore suggested to look for a factorization which shows the
entanglement most clearly.\\
\\
3. The change from a factorization which shows interaction / entanglement
into a factorization without interactions is applied in one very basic example
of QM: the hydrogen atom. (The connection with our more global factorization
problem has been pointed out already in \cite{dugic1}.) The proton and 
electron interact with each other and make the hydrogen atom a two-body problem.
It is, however, possible to disentangle the system into two 
independent one-body problems:
one free system, given by a wave function of the center of mass position,
$\psi_\mathrm{cm}(\mathbf{x}_\mathrm{cm},t)$,
and one system given by a wave function of the relative position of proton and 
electron, $\psi_\mathrm{rel}(\mathbf{x}_\mathrm{rel},t)$,
residing in a Coulomb potential. 
There is no interaction term between these two wave functions. This example
is analogous to the disentanglement of the measurement process demonstrated
in this section.

\section{Conclusions}
I have shown that it is always possible to factorize the global 
Hilbert space into subsystems in such a way, that the story told
by this factorization is that of a world in which nothing happens.
A factorization into {\it interacting} and
{\it entangling} subsystems is also possible, in infinitely many
arbitrary ways. But such a more complicated factorization is meaningful
only if it is justified through interactions with an external observer
who does {\it not} arise as a part of the state vector.

The Many World Interpretation is therefore rather a No World Interpretation
(according to the simple factorization), or a Many Many Worlds Interpretation
(because each of the arbitrary 
more complicated factorizations tells a different story
about Many Worlds \cite{dugic2}).

We can say that the state vector of QM 
{\it per se} does not contain any information or substructure.
The state vectors of norm 1 are all equivalent, because the scalar
product is the only internal structure of a Hilbert space.
Information arises only when a preferred basis 
is somehow singled out, according to which
the state of the system can be written in components. 
(A part of the choice of basis is the choice of factorization, the split of
the system into subsystems, which gives it an internal structure.)
The information
is then contained in the relation of the state to the basis, not in the
state alone. The notion of information or structure
which arises from this view
is interesting and requires further discussion. Information or structure
in this sense
is not contained in an object, but only in the object's relation to
a receptor, or more precisely: the object's relation to the way in which
the receptor looks at the object. Another example in a very different context 
is discussed in section 6.2.

The state vector of the universe in the EI
has no environment or observer it can relate
to, and is therefore completely meaningless. 
The appearance of interacting subsystems 
of the universe are only due to a choice
of a ``samsara" basis, which is however completely arbitrary, just like
a slicing of Minkowski spacetime is possible, which makes it look
like an expanding universe (cf eq.\ref{minksams}).

One has to add something to give the state vector and QM a meaning.
First one can mention several things that will {\it not} help to solve
the factorization problem:\\
\\
1. Dynamical Collapse \cite{grw}: This is a modification of the Schr\"odinger
equation to single out one of the ``many worlds" dynamically.
This has nothing to do with our concerns here and is not helpful at all.
It leads only to a change of the time evolution of the state vector,
which can again be absorbed by a comoving basis. The factorization
problem is unaffected.\\
2. Postulation of spacetime: If spacetime is postulated as an additional
entity, it becomes ``natural" to construct the Hamilton operator from
operators acting locally in space. But an observer would still have
to arise from a substructure of the state vector (we didn't postulate
any additional objects in spacetime, only spacetime itself), and here is the
problem: As discussed in sections 2 and 3, the state vector doesn't
``see" the space that went into the construction of the Hamilton operator.
The factorization problem remains. \\
3. Dualistic
interpretations of QM which emphasize the role of ``minds",
like ``Many Minds" \cite{manyminds1,manyminds2}, 
or like interpretations which make mind responsible
for a collapse of the state vector \cite{wigner,stapp}. 
In this case it is the burden of the
minds to see the world according to a certain factorization. But since
the state vector does not contain {\it any structure at all}, such an
approach shifts all structure {\it completely} into the minds, or their
reception of / interaction with
the state vector, or the way in which the state vector
guides the minds, or whatever is the appropriate terminology in a
specific interpretation. I.e. all structure is pushed outside of
what is in reach for scientific analysis. This cannot be regarded a
meaningful solution.\\

There are several possibilities to avoid the factorization problem:
\begin{itemize}
\item {\bf Pilot wave theory} (PWT), also known as Bohm theory
 \cite{bohm,holland}: 
 Particles moving in spacetime are added
 to the theory. I.e. a spacetime is postulated, and in addition
 particles moving in this spacetime are postulated. ``Observers"
 consist of these particles, not of substructures of the state vector.
 The particles are ``guided" by the state vector. 
 Pilot wave theory has been criticized by EI proponents as unnecessary: 
 The many worlds of the EI are also present here within the state vector,
 and the particles serve only as ``pointers" to one of these worlds
 \cite{antibohm}.\\
 But now we see that PWT solves the main problems discussed here:
 The particles define a meaningful factorization of the global
 Hilbert space (they are by definition the ``objects", i.e. subsystems
 of the universe), 
 and a position space that serves as a preferred basis.
 The state vector is defined {\it as} a wave function in a configuration
 space corresponding to particle positions. 
 The feature of the PWT, that it brings a preferred basis into the theory, is not 
 a weakness, it is a strength.
 It remains to be seen, however, how well PWT can be adapted to a quantum theory
 as complicated as the Standard Model of Particle Physics.    
\item {\bf Copenhagen Interpretation} (CI): 
 The CI is actually a statement about science.
 The world as far as it can be described
 by science must be split into a realm of phenomena and a realm of modeling.
 The realm of phenomena contains subjects making observations on objects,
 and the objects are described in a ``classical way"
 (on a first layer of modeling), i.e. with 
 definite properties, and localized in spacetime. 
 The realm of modeling (actually: the second layer of modeling)
 is a mathematical structure (here: a Hilbert space) whose only purpose is
 to make predictions about the phenomena. It does not have an independent
 existence, and therefore there is no factorization problem.
 The question why there is a certain relation between the model and the 
 phenomena, i.e. why the model actually predicts the phenomena, 
 cannot be solved by science. 
 The CI was often criticized as unclear and anti-realist. But one can 
 also see it as a positive way to deal with the
 limitations of science. As an approach which
 accepts these limitations, and which makes the distinction between 
 a model to describe phenomena and a model to predict phenomena
 (or their statistics) explicit;
 More comments on the CI and its relation to the EI follow in section 6.4.
\item {\bf QM as an effective theory}: QM is only a limit of a
 more fundamental theory. The state vector is therefore only an
 emergent object, not a fundamental one \cite{cw}.
 The appearance of interacting objects must be explained in 
 the fundamental theory, not from the state vector.   
\end{itemize}

\section{Connections to other topics and Outlook}
\subsection{Epistemic versus ontic state}
There is an ongoing discussion whether the state vector has ``ontic" or
``epistemic" character, where, roughly speaking, ``ontic" means it is
a real thing ``out there", and ``epistemic" means it is just an effective
way to charactarize an observer's 
subjective state of knowledge about a system.

One year ago, Pusey et al published a paper \cite{pusey} which
showed that certain interpretations treating the state vector
as epistemic are inconsistent. 
(In particular, they showed that it is not possible to have a ``real
physical state" that is consistent with several quantum states, such
that each of these quantum states can be interpreted as 
representing some subjective
knowledge about the real state.)
Here I have shown that the state vector also 
cannot be ontic, if ``ontic" implies that it can stand by itself,
without a relation to other objects or observers.

The strange hybrid character of the state vector comes from the fact
that it makes sense only in relation to a measurement. On the other hand,
it can describe the measurement process as part of itself, but only
if the result is subjected to another measurement
(as shown in section 4). I will come back
to the strange double role of measurement in section 6.4. 

In my opinion, the whole notion of ontology and its relation to 
science should be rediscussed with a fresh attitude.
I will present some ideas in that direction in a separate paper.

\subsection{Story of a Brain}
An argument very similar to the one presented here has been given by 
Zuboff in his {\it Story of a brain} \cite{dh}
to reject certain theories about how
subjective experience arises from brain states. These theories contained
certain assumptions about which aspects of the brain states are irrelevant
to the generated experience. From these assumptions Zuboff constructed
transformations of the geometry and sequence of the
firing / non-firing neurons, which should leave the experience unaffected.
He was able to combine the transformations in such a way that finally
the ``brain" consisted of only one neuron which fires one single time.
He concluded that if these theories were right, every possible subjective
experience would be contained already in a single firing neuron. Hence
these theories are inconsistent.

The analogy with the present argument is obvious. Zuboff's transformations
correspond to the basis transformations in QM. Then it turns out there is 
a basis in which the state appears trivial, i.e. strucutureless. Therefore, 
the state {\it is} trivial and cannot explain 
(or is in contradiction with) the complicated patterns seen in the universe / 
in subjective experience.

It is instructive to see how Hofstadter and Dennett try to refute Zuboff's
argument (they are proponents of these theories he has taken {\it ad absurdum}):
They argue, if Zuboff's argument were right, one could also say that the
entire information contained in all the books of the world are
already contained in printing each letter from A to Z one single time. 

But here is exactly the difference I have discussed in the context of QM:
The books in the libraries are read by {\it external observers}. These
observers read the books according to a specific basis: They read every line
from left to right, each page from top to bottom etc. The books
contain their information {\it with respect to that ``basis"}, and therefore
printing each letter a single time is insufficient, because the corresponding
``basis transformation" is not allowed. The flickering of the neurons, on 
the other hand, should lead to subjective experience {\it per se}, without
the help of an external observer, and therefore without a preferred ``basis".
Therefore, Zuboff is right, and Dennett and Hofstadter are wrong.

\subsection{Is the world a mathematical structure?}
Tegmark has argued convincingly that science reduces the world
to a mathematical structure \cite{tegmark1,tegmark2,tegmark3}.
I.e., the picture of the world that is created by science {\it is}
a mathematical structure.
In particular, in science it doesn't make sense to make a distinction
between the structure and entities described by the structure,
because there is nothing that science could say about such entities except for
what is already contained in the structure.

Tegmark's view is clearly that this picture created by science is complete,
i.e. the mathematical strucuture can stand on its own ground and {\it is}
the world (or the multiverse) in which we live and are a part of.

This view is challenged by the argument I have given here. 
If we take QM in the EI,
the mathematical structure is the state vector and its time evolution.
This structure has actually turned out to be empty if it stands on its own.
It {\it becomes} a 
nontrivial structure only in relation to an external observer, or
through the interaction with an environment which is not part of the state
vector already.

PWT doesn't seem to have this problem at first glance. The structure is now 
given by the state vector {\it and} the particle trajectories. But if
we take the mathematical universe serious, 
on a fundamental level we shouldn't even speak of particles as entities
separate from the mathematical structure. There are {\it only the trajectories},
and it is a pure language convention to assign them to entities called 
particles. So, the trajectories form a set of $3n$ real-valued
functions of time (where $n$ is the number of particles). 
What kinds of transformations should we allow on the $3n$ functions and
claim that these transformations don't change the structure,
or don't change the ``information" contained in the structure? Why 
should we consider the $3n$ functions as trajectories of $n$ particles
in 3 dimensions and not, say, of 3 particles in $n$ dimensions?
I.e. how is our observation of a 3-dimensional space filled with
objects reflected in the ``structure" of trajectories?   
It seems to me that we run into problems similar to the EI, and that the 
following statement is fundamental:\\
\\
A structure is a structure only with respect to some 
observer or environment outside the structure who reads off the 
structure {\it as} a structure in a specific way.\\
\\
This statement requires further discussion and justification.
At the moment one can say that 
the only interpretation of QM which has this statement built in is the CI.

\subsection{Schr\"odinger's Problem}
In \cite{erwin1}, Schr\"odinger writes:\\
 
``The thing that bewilders us is the curious double role that the conscious mind 
acquires. On the one hand it is the stage, and the only stage on which this whole
world-process takes place, or the vessel or container that contains it all and
outside which there is nothing. On the other hand we gather the impression, maybe
the deceptive impression, that within this world-bustle the conscious mind
is tied up with certain very particular organs (brains), which [...]
serve after all only to maintain 
the lives of their owners, and it is only to this that they owe 
their having been
elaborated in the process of speciation by natural selection."  \\ 

Schr\"odinger compares the situation with an artist who places a 
picture of himself
as an inconsiderable minor character in one of his paintings. This seems to him
the best allegory for the confusing double role of mind. 
On the one hand it is the
artist who created everything; in the completed work, however, 
it is only an unimportant
decoration which could have been left out without changing the total impression
substantially.\\

The text is part of Schr\"odinger's philosophical work, at that time totally
unrelated to QM. But now we see that the situation in QM is similar,
if we replace ``mind" by ``measurement": In the CI, the measurement is
central. It is the only stage on which the whole physics-bustle takes place,
and the state vector is justified only as a model to predict the statistics
of measurement results, with no independent existence. In the EI, 
the measurement process is just a little interaction process like many others,
and is completely included in the picture of the state vector, as a
``minor character" that could be missing without changing the picture 
substantially. This is the strange double role of the measurement.

It seems that one of these views alone cannot survive. If the CI is taken
alone, one may respond: ``But why should we mystify the measurement?
I can model the measurement inside QM, as a part of the unitary evolution
of the state vector, without giving it such a fundamental role." 
If, on the other hand, the EI is taken alone, we have seen that the 
resulting picture is not a picture anymore. It is an empty nothing.
Only together, as complementary views on QM, the CI and EI make sense.

The strange double role of the measurement, just as the strange double role
of the mind, is a problem most fundamentally related to what we do
when we do science. We create a picture of objects; a picture created
by subjects. The double role is fundamentally built into science.
I conjecture it cannot be resolved within science.

\end{document}